\documentclass[twocolumn,prb,superscriptaddress,amsmath,amssymb,aps]{revtex4}
\usepackage{amsmath}
\usepackage{bm}
\usepackage{braket}
\usepackage{graphicx,subfigure}
\usepackage{color}
\usepackage{url}
\usepackage[latin1]{inputenc}
\usepackage{tikz}
\usepackage[toc,page]{appendix}
\usepackage[normalem]{ulem}

\makeatletter
\def\BState{\State\hskip-\ALG@thistlm}
\makeatother

\begin{document}

\title{Off-stoichiometric softening and polytypic transformations in the plastic deformation of the C14 Fe$_2$Nb Laves phase}
\author{A.N.~Ladines}
\affiliation{Atomistic Modelling and Simulation, ICAMS, Ruhr-Universit\"at Bochum, D-44801 Bochum, Germany}
\author{R.~Drautz}
\affiliation{Atomistic Modelling and Simulation, ICAMS, Ruhr-Universit\"at Bochum, D-44801 Bochum, Germany}
\author{T.~Hammerschmidt}
\affiliation{Atomistic Modelling and Simulation, ICAMS, Ruhr-Universit\"at Bochum, D-44801 Bochum, Germany}

\begin{abstract}
Plastic deformation of the brittle C14-Fe$_2$Nb Laves phase occurs mostly by basal slip due to the complex crystal structure.
Here, we compare the barriers for basal slip for the known mechanisms of direct slip, synchroshear and undulating slip using density functional theory calculations.
According to our calculated generalized stacking fault (SF) energies, the most favorable mechanisms are synchroshear and undulating slip.
Both mechanisms lead to stable SF with a formation energy of 50~$mJ/m^2$ through the same unstable SF configuration at the transition.
The energy barrier of approximately 3~$J/m^2$ indicates a low dislocation mobility as expected from the brittle character. 
We also determine the influence of vacancies and antisite defects on the formation energy of stable and unstable SF. 
Both kinds of point defects tend to lower the energy barrier on both sides of 2:1 stoichiometry.
This explains the experimentally observed off-stoichiometric softening of C14-Fe$_2$Nb. 
The small energy differences between the Fe$_2$Nb Laves phase polytypes raises the question if there are further deformation mechanisms with low barrier.
Therefore, we additionally consider transformations between C14, C15 and C36 Laves phases as further deformation mechanism. 
Our calculations for polytypic transformations by successive synchroshear steps show that the corresponding energy barriers are in fact very similar to the energy barrier for basal slip in C14.
This suggests that the energy needed to create a stable SF in C14 by synchroshear is also sufficient to initiate polytypic transformations where existing SFs in C14 are further transformed to form C15 or C36 Laves phases. 
\end{abstract}

\maketitle

\section{Introduction}

One of the most commonly observed group of crystal structures are the topologically close-packed (TCP) phases~\cite{Sinha1978} and particularly the Laves phases C14, C15, C36. 
Laves phases have been subject to extensive theoretical and experimental studies, see e.g. Ref.~\onlinecite{Stein2021} for a recent review. Their structural stability is to a large degree governed by differences in atomic size and the average number of valence electrons~\cite{Seiser-2011-1,Seiser-2011-2,Hammerschmidt2013,Ladines2015}.
The Laves phases are inherently brittle at low temperatures($<0.6T_m$)~\cite{Livingston1992}, a result of the complex crystal structure of the Laves phases that lacks straightforward slip mechanisms~\cite{Heggen2010,Zhang2020}.

\begin{figure}[htb]
 \centering
 \subfigure[]{ \label{fig:laves_quadruple}         
 \includegraphics[width=0.8\columnwidth]{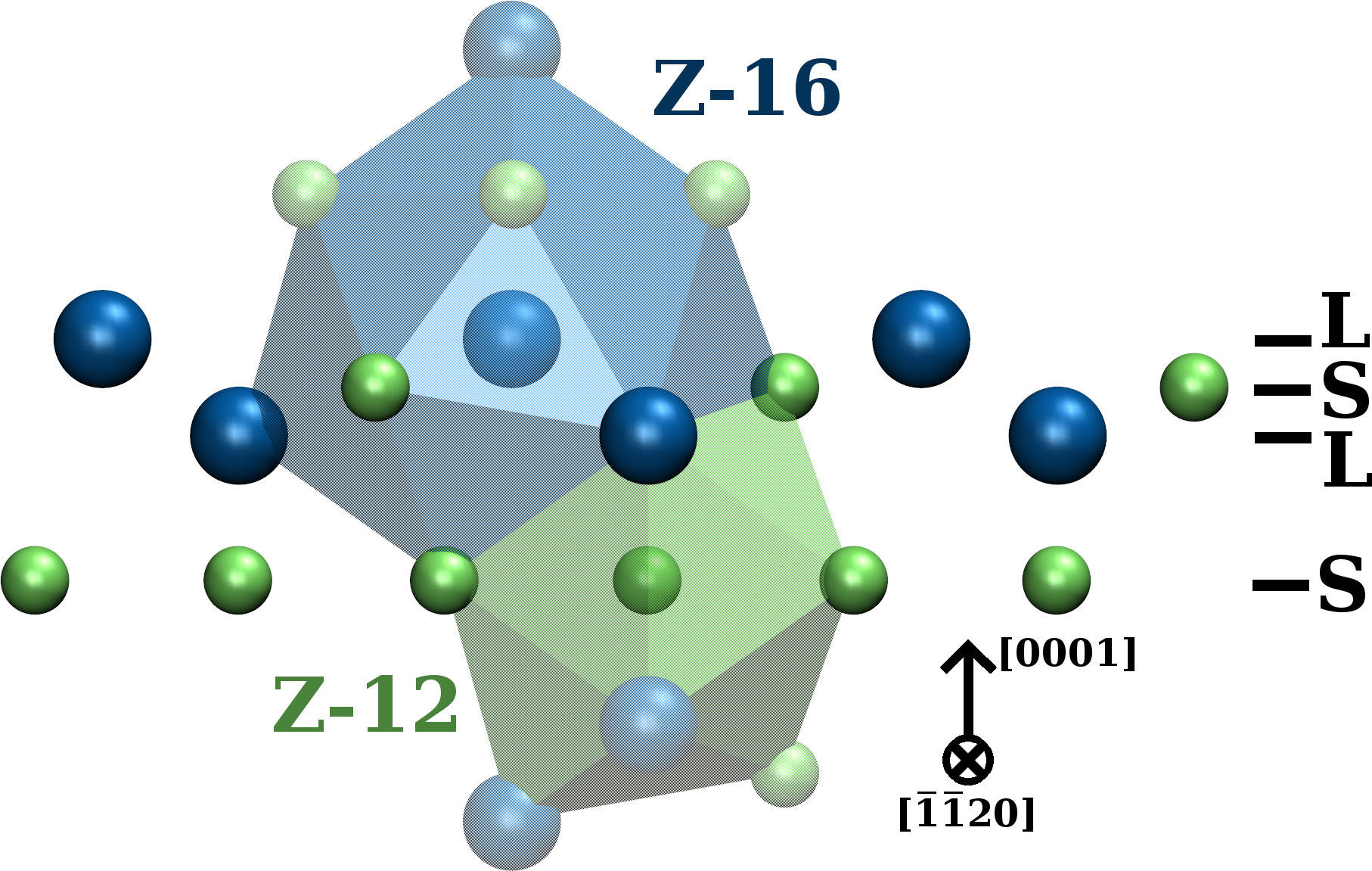}
}
\subfigure[]{ \label{fig:laves_sites}         
 \centering
 \includegraphics[width=0.8\columnwidth]{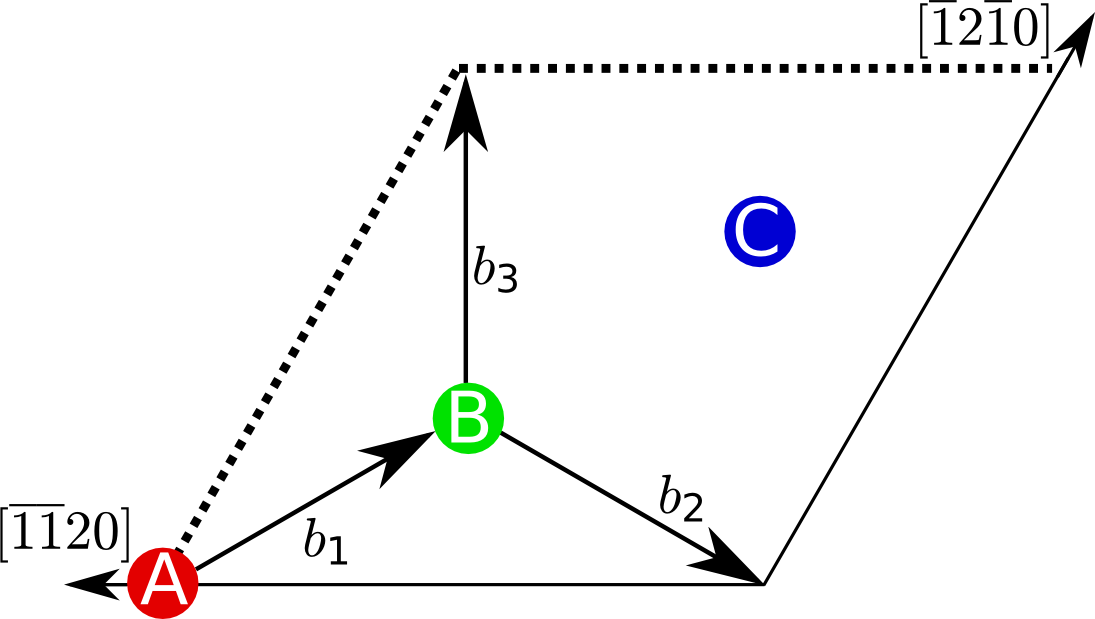}
 \label{laves_sites}
 }
 \caption{(a) Stacking sequence of the elementary quadruple layer in Laves phases viewed parallel to the basal plane. 
          The Kagom\'e layer contains smaller atom only while the triple layer is a stacking of large-small-large atoms. 
          (b) High symmetry points and Burgers vectors in the hexagonal lattice representation of Laves phases.}
 \end{figure}

The crystal structure of the Laves phases comprises two-third polyhedra with coordination number $Z$=12 and one-third $Z$=16 polyhedra, occupied preferably by small (S) and large (L) atoms, respectively. 
It consists of a close-packed stacking of atomic layers of the $6\cdot 3\cdot 6 \cdot 3$-type in the Sch\"afli notation known as Kagom\'e layer and a puckered triple layer of L-S-L atoms along the $\langle$0001$\rangle$ direction in the hexagonal (h) representation (see Fig.~\ref{fig:laves_quadruple}) or along the $\langle$111$\rangle$ direction in the cubic (c) representation.
The interlayer spacing within the puckered layer is approximately one third of that between the triple layer and the Kagom\'e layer. 
The atomic layers can be centered on either of the three high-symmetry points which are related by the Shockley Burgers vectors $b_1 = a/3 \langle 1\bar 100\rangle$-h or $b_2 = a/6{\langle}112\rangle$-c as shown in Fig.~\ref{fig:laves_sites}. 
In the C14-Fe$_2$Nb Laves phase, the Fe atoms correspond to S atoms on 2a and 6h sites with $Z$=12 and the Nb atoms correspond to L atoms on 4f sites with $Z$=16.

The stacking sequence needs to fulfill certain requirements in order to achieve close packing: 
the lower part of the triple layer occupies the center of the Kagom\'e layer while the middle sublayer occupies one of the two other high symmetry points. 
This results in two variants of the triple layer, $t$ and its reflection $t'$. 
The top layer is centered on the remaining high symmetry point. 
Depending on the centering of the Kagom\'e layer, the quadruple layer can be of type $X$, $Y$, $Z$ or their reflections $X'$, $Y'$, $Z'$. 
The $X$ quadruple layer, e.g., corresponds to an A$\alpha$c$\beta$ stacking sequence.
(Small atoms on the A, B and C points are denoted as a, b, and c while large atoms on these sites are referred to as $\alpha$, $\beta$ and $\gamma$.)
The various stacking sequences of quadruple layers give rise to the Laves phase polytypes C14, C15 and C36 (see Ref.~\onlinecite{Kumar2004} for a detailed discussion) with the three shortest repeating periods $XY'$ (h), $XYZ$ (c) and $XY'X'Y$ (h). 

The experimentally observable slip in Laves phases occurs mostly in the basal plane~\cite{Allen1972,Kumar1994,Chisholm2005}. 
The three known basal deformation mechanisms for Laves phases are direct slip, synchroshear ~\cite{Hazzledine1993} and undulating slip~\cite{Zhang2011}. 
Their relative importance is determined by the generalized stacking fault energy (SFE) curve~\cite{Tadmor2003,VanSwygenhoven2004}. 
While hardly accessible by experiment, the generalized SFE can be determined by atomistic calculations using electronic-structure methods~\cite{Moeller2018}.
For several Laves phase compounds, the generalized SFE has been determined by density-functional theory (DFT) calculations~\cite{Vedmedenko2008,Ma2013,Ma2014}.
For the technologically important C14-Fe$_2$Nb Laves phase~\cite{Voss-11}, however, atomistic insight to the governing mechanism of plastic deformation is still missing.

In this study, we perform DFT calculations to determine the generalized SFE curves in the C14-Fe$_2$Nb Laves phase. 
We compare the deformation by direct slip, synchroshear and undulating slip in defect-free, stoichiometric C14-Fe$_2$Nb, followed by an analysis of the influence of off-stoichiometric compositions due to vacancy or antisite defects. In addition we assess the possibility of polytypic transformations from C14 to C15 and C36 by successive transformation via intermediate SF configurations.

\section{Methodology}
\label{sec:methodology}

The DFT calculations presented in this work were performed with the VASP software (version 5.4)~\cite{Kresse1996,Kresse1996b,Kresse1999} using the projector-augmented wave method~\cite{Bloechl-94}. 
The generalized gradient approximation to the exchange-correlation functional~\cite{Perdew-96} and pseudo-potentials with $p$ semicore states were employed. 
A planewave cut-off energy of 450~eV and a $k$-point spacing of 6~\AA$^{-1}$  was required to converge the total energies to less than 1~meV/atom. 
A supercell consisting of 24 atomic layers (36 atoms) was used in the calculation of the bulk SFE. 
The initial magnetic configuration was set to the energetically most favorable arrangement for the Fe$_2$Nb C14 Laves phase, i.e. (UD)/(UD)/U on the 2a/6h/4f sites, as determined previously~\cite{Ladines2015}. 
No constraints on the spins were applied during the electronic self-consistency. 
The reference structure was obtained by optimizing the lattice parameters and the atomic positions of the bulk C14-Fe$_2$Nb Laves phase. 
For the deformation along the generalized SFE curve we use sheared periodic boundary conditions that vary as a function of the deformation parameter~\cite{Vedmedenko2008}.
During the deformation, the ionic positions were allowed to relax while the deformed simulation cell was kept fixed. 
For the simulations with vacancy and antisite defects, a $2\times2\times2$ supercell (96 atoms) was used to avoid interaction of the points defects with their periodic boundary image. 

\section{Results and discussion}

\subsection{Basal deformation of stoichiometric C14-Fe$_2$Nb}
\label{sec:SFE}

\subsubsection{Deformation mechanisms}

\begin{figure}[htb]
 \centering
 \subfigure[]{
 \includegraphics[width=0.3\columnwidth]{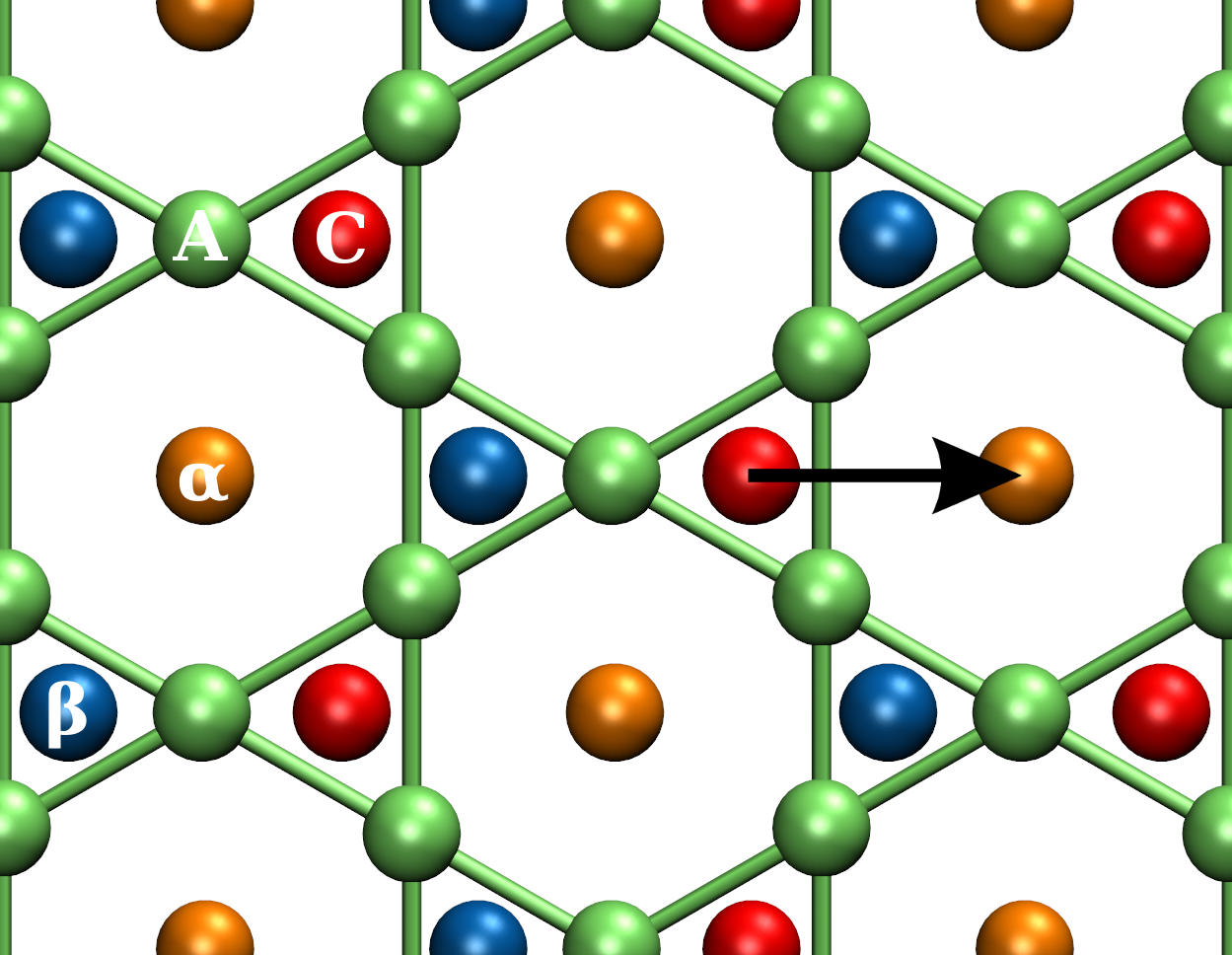}}
 \subfigure[]{
 \includegraphics[width=0.3\columnwidth]{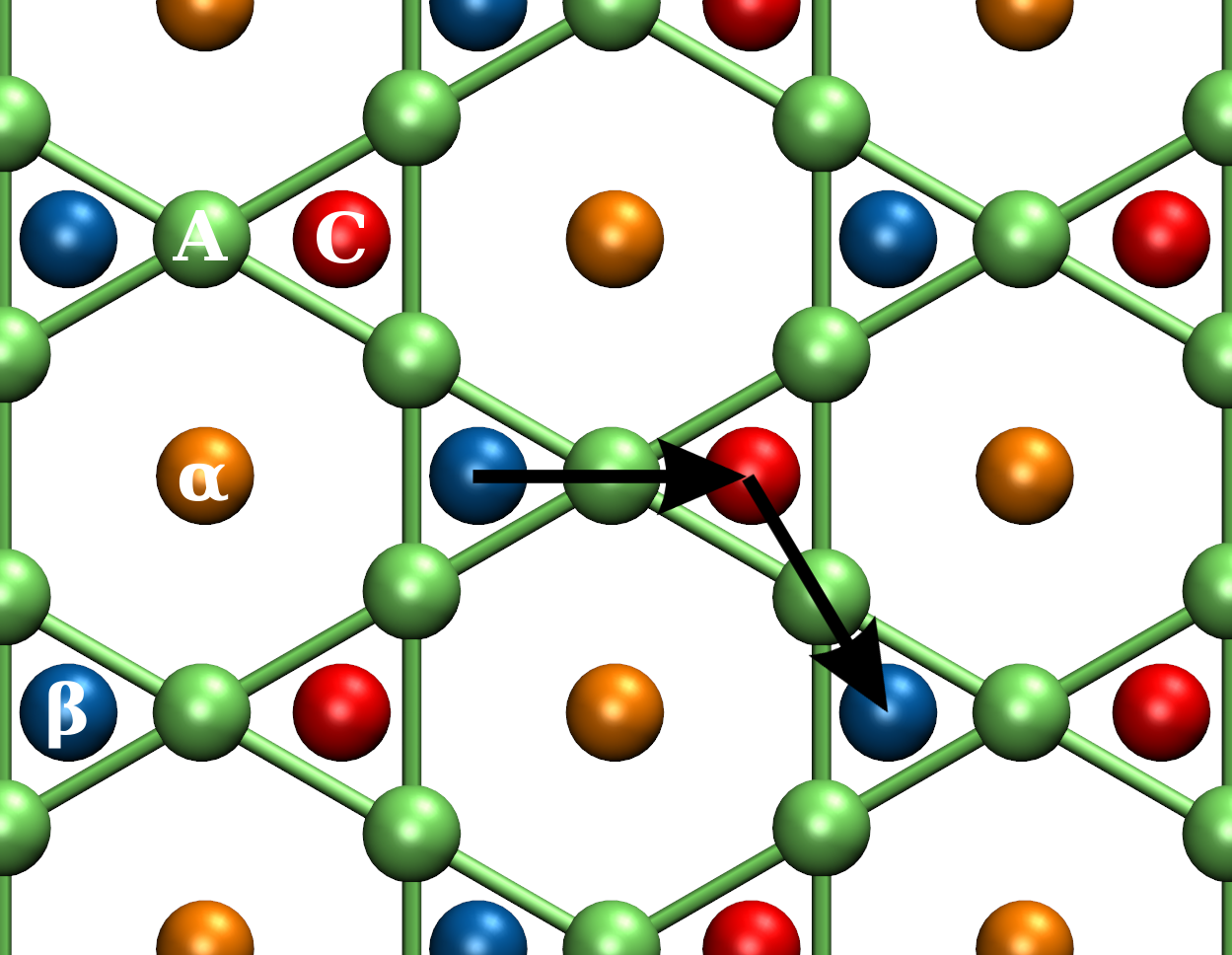}}
 \subfigure[]{
 \includegraphics[width=0.3\columnwidth]{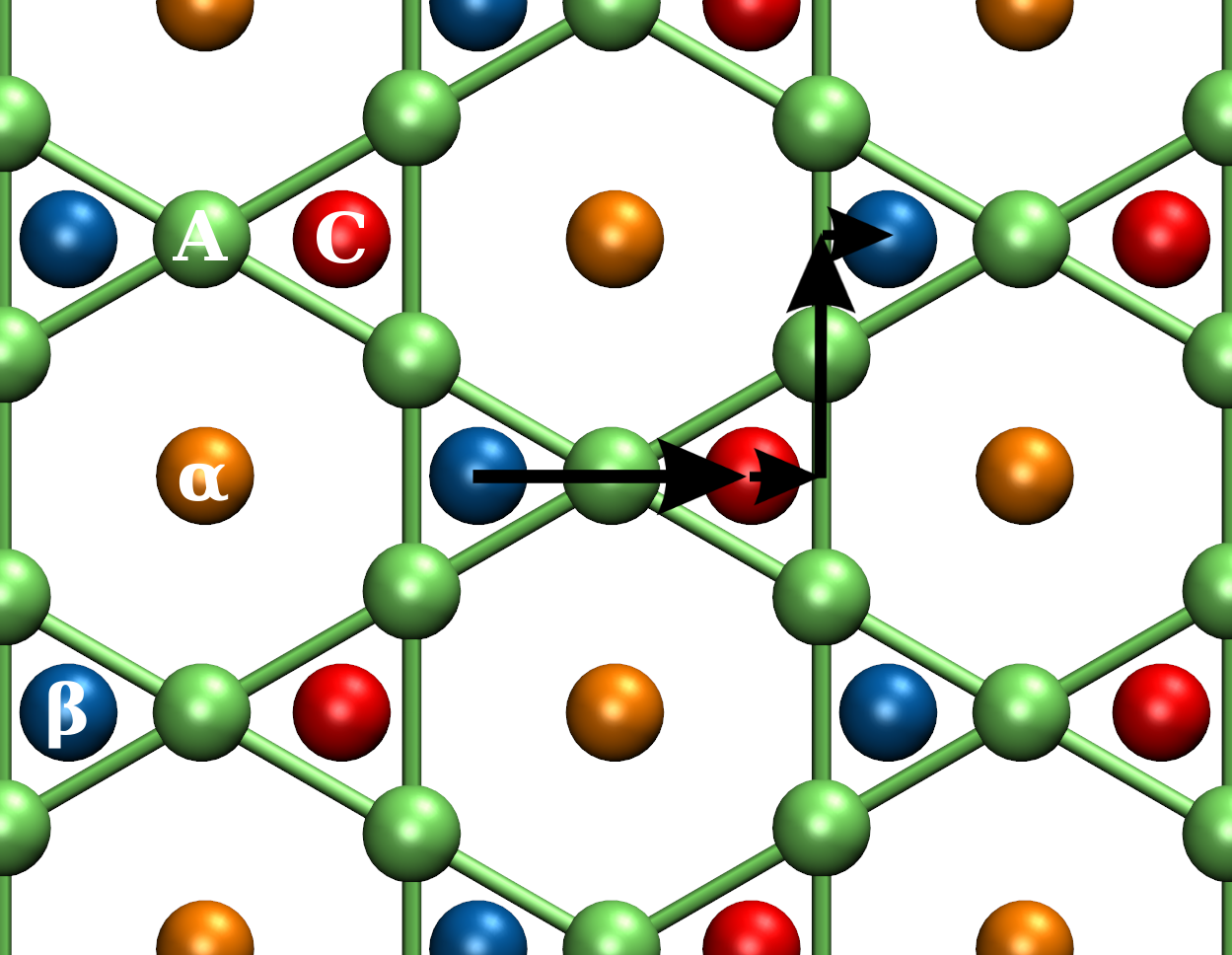}}\\
 \subfigure[]{
 \includegraphics[width=0.3\columnwidth]{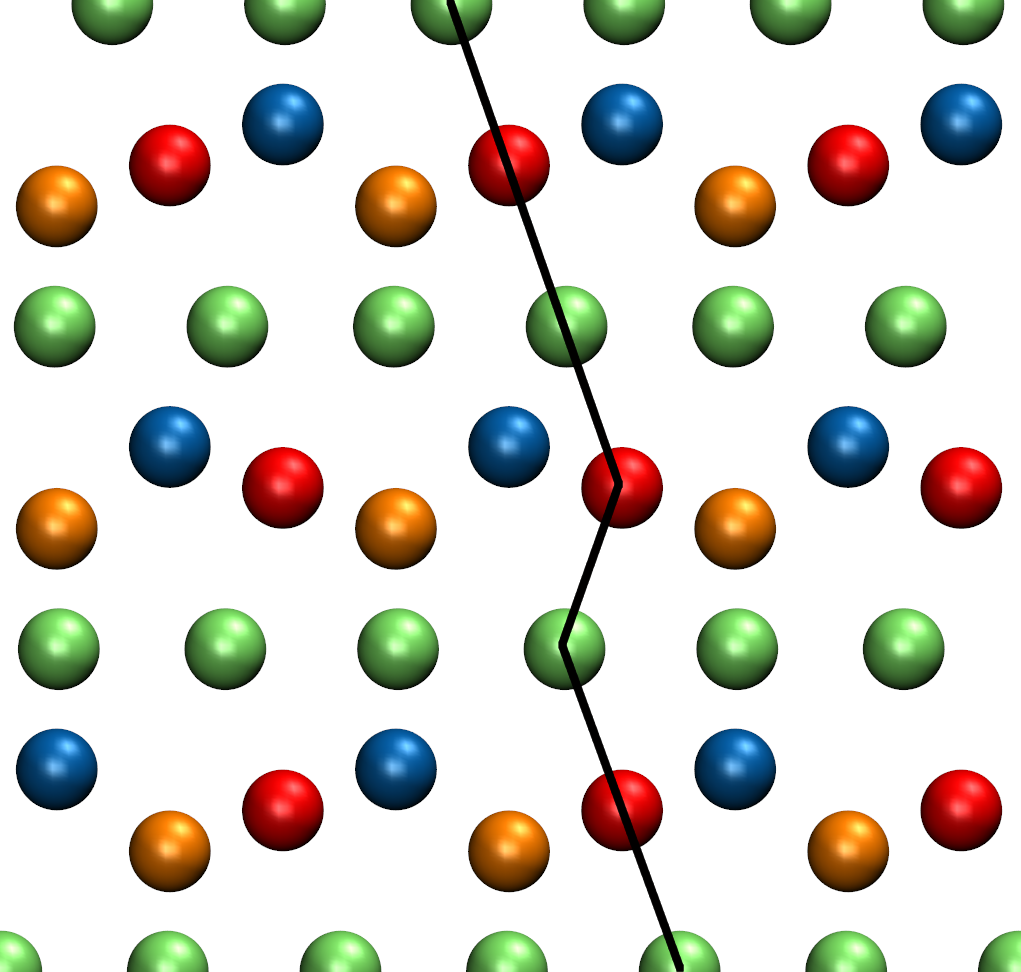}}
 \subfigure[]{
 \includegraphics[width=0.3\columnwidth]{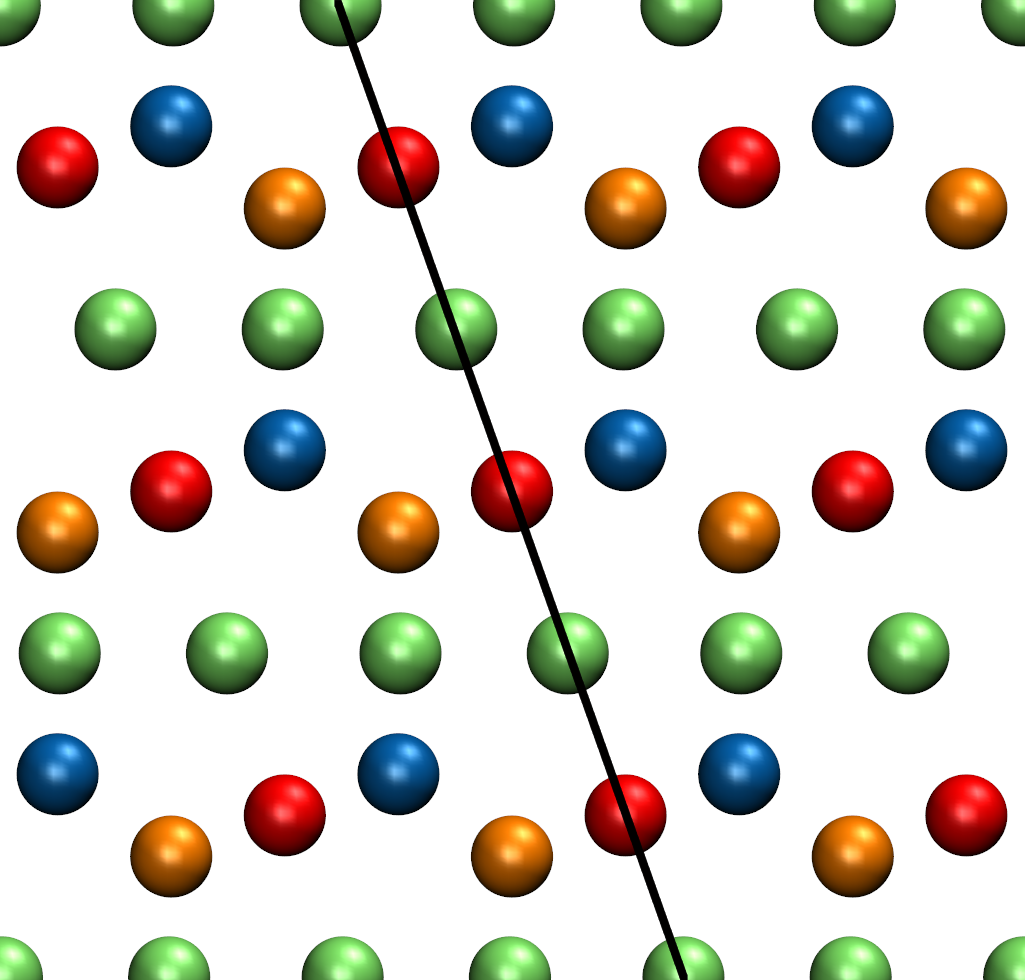}}
 \subfigure[]{
 \includegraphics[width=0.3\columnwidth]{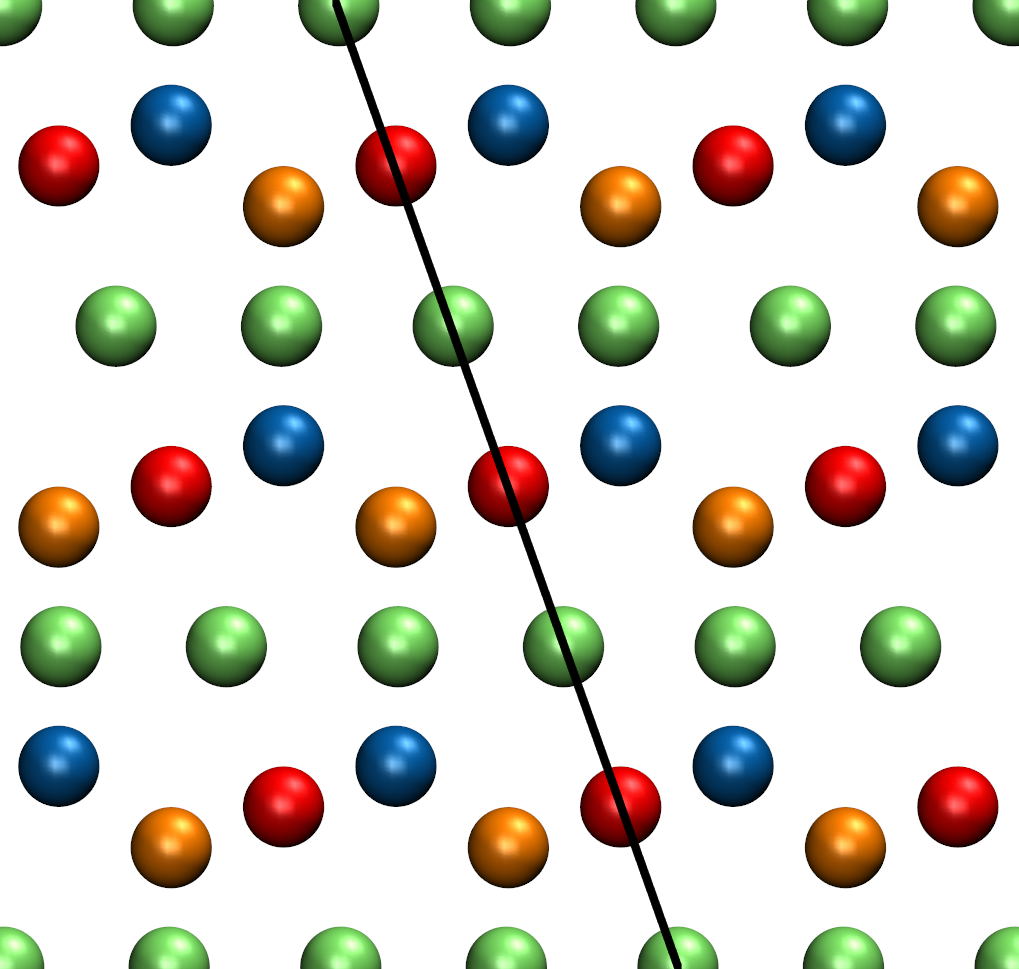}} 
 \caption{Basal deformation modes in Laves phase: direct slip (left), synchroshear (middle) and undulating slip (right)
          viewed along $[0001]$ (Figs. a-c) and $[\bar{1}\bar{1}20]$ (d-f). 
          The arrows in the upper figures indicate the motion in the quadruple layer.
          The upper half of the crystal above the deformation plane is also shifted in the slip direction
          by one Burgers vector.  The solid lines in the lower figures indicate the layer sequence in the 
          resulting SF.}
 \label{fig:laves_deform}         
\end{figure}

There are at least three basal deformation modes in Laves phases, shown schematically in Fig.~\ref{fig:laves_deform}: 
(i) The most apparent mechanism is direct slip between the Kagom\'e and the triple layer with Burgers vector $a/3<10\bar 10>$-h.  
(ii) The synchroshear slip mechanism involves a synchronous displacement of the middle and top layer by $a/3<10\bar 10>$-h and $a/3<01\bar 10>$-h, respectively~\cite{Hazzledine1993}.  
(iii) The undulating slip mechanism is a deformation process which involves crystallographic slip and atomic shuffling.
This slip mechanism was shown to have a lower energy barrier in the Cr$_2$Nb system compared to the previously discussed deformation modes~\cite{Zhang2011}.

\subsubsection{Generalized SFE curves for C14-Fe$_2$Nb}

\begin{figure}[htb]
 \centering
 \includegraphics[width=0.95\columnwidth]{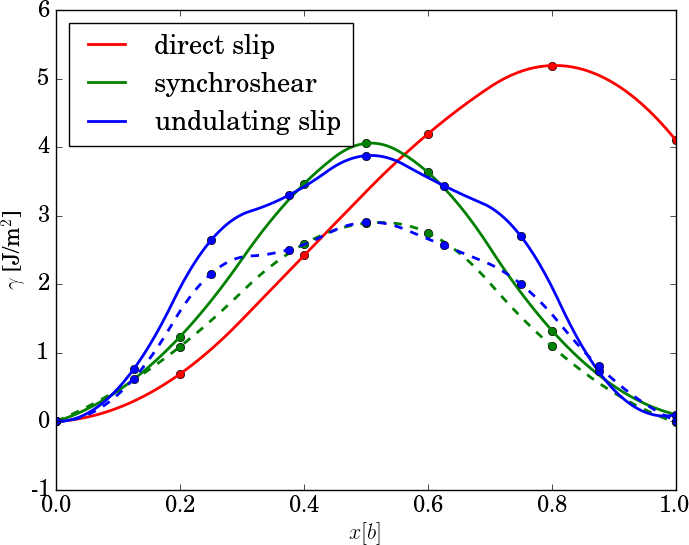}
 \caption{Generalized SFE curve corresponding to the basal deformation modes in C14-Fe$_2$Nb. 
          The points denote the calculated SFE while the lines are second-order spline interpolations. 
          The dashed lines connect the stacking fault energies in the relaxed configuration.}
 \label{fig:SFE}         
\end{figure}

Experimental analysis of deformed Fe$_2$Nb samples identified SF structures as remainders of basal slip deformation~\cite{Takata2008,Slapakova2016} and suggested synchroshear as dominant deformation process.
However, undulating slip cannot be excluded as the resulting local C15 stacking sequence is the same for synchroshear and undulating slip (see Fig.~\ref{fig:laves_deform}). 
Here, we determine the most favorable basal deformation mechanism in C14-Fe$_2$Nb by DFT calculations of the energy barrier along the generalized SFE curves shown in Fig.~\ref{fig:SFE}. 

Direct slip involves rather small energies in the initial stage due to the large free volume between the Kagom\'e layer and the triple layer which then increase as the Fe atoms in the Kagom\'e layer approach the Nb atoms in the adjacent layer. 
The energy barrier for direct slip, given by the maximum of the generalized SFE curve, is the largest among the three deformation mechanisms.
This deformation results in a metastable configuration which may further develop to a stable configuration through a similar slip along the two other Burgers vectors. 
Synchroshear and undulating slip, in contrast, lead to a stable SF with a formation energy of $\approx$50 mJ/m$^2$. 
The energy barrier to arrive at this stable SF is reported in Fig.~\ref{fig:SFE} for both, relaxed and unrelaxed configurations. 
For the unrelaxed configurations, the undulating slip has the lowest energy barrier of the three deformation modes.
Upon relaxation, the synchroshear and undulating slip lead to the same unstable SF configuration at the maximum of the generalized SFE curve and hence exhibit the same energy barrier of approximately 3~J/m$^2$.
This finding is in line with the experimentally observed SF configuration after deformation~\cite{Takata2008,Slapakova2016}.
Similar results were obtained recently with a classical potential for the case of C14-Mg$_2$Ca~\cite{Guenole-19}.

Overall, the rather large barriers indicate that plastic deformation of the C14-Fe$_2$Nb Laves phase by dislocation nucleation in the basal plane is associated with large stresses. 
This raises the questions (i) if the barriers can be reduced, e.g., by changing the chemical composition (see Sec.~\ref{sec:defects}), (ii) if further processes are activated by the required large stresses (see Sec.~\ref{sec:polytypic}) and (iii) if the experimentally observed SF in C14-Fe$_2$Nb could be remains of other, not yet identified processes.

\subsection{Off-stoichiometric softening}
\label{sec:defects}

\subsubsection{Formation energy of point defects in C14-Fe$_2$Nb}

Many Laves phases are known to form also at off-stoichiometric compositions~\cite{Thoma1995}.
The deviation from the 2:1 composition is attributed to vacancies or antisite atoms, primarily of the smaller atom. These point defects are known to have profound impact on the mechanical properties~\cite{Liu2000}. 
The vacancy formation-energy
\begin{equation}\label{eq:dH1}
    E_f^{\mathrm{vacancy}} = \frac{(1-x_D) \Delta H_{\mathrm{vacancy}}-\Delta H_{\mathrm{ideal}}}{x_D}
\end{equation}
and the antisite formation energy 
\begin{equation}\label{eq:dH2}
    E_f^{\mathrm{antisite}} = \frac{\Delta H_{\mathrm{antisite}}-\Delta H_{\mathrm{ideal}}}{x_D}
\end{equation}
are computed from the difference in the heat of formation obtained by DFT as outlined in the Appendix. $x_D$ is the fraction of defects to the number of atoms in the supercell. The heat of formation of the supercells with and without defect are computed with respect to the bcc ground-states of Fe and Nb. (A previously reported simpler accounting for changes in chemical composition due to defect formation leads to qualitatively comparable results~\cite{Ladines-17}.)

\begin{table}[htb]
 \centering
 \caption{Vacancy and antisite formation-energies in C14-Fe$_2$Nb.}
 \scalebox{1.00}{
 \begin{tabular}{l c c }
 \hline
 \hline
 site           & $E_f^{\mathrm{vac}}$(eV)   & $E_f^{\mathrm{antisite}}$(eV) \\
 \hline  
 2a (Fe)      & 2.71    & 0.60   \\
 6h (Fe)      & 2.57    & 0.87   \\
 4f (Nb)      & 2.77    & 1.33   \\
 \hline
 \hline
 \end{tabular}}
 \label{tab:ef_vac_anti}
\end{table}

Our DFT calculations, compiled in Tab.~\ref{tab:ef_vac_anti}, show that the formation energies of the different antisite defects are considerably lower than all vacancy formation energies. We therefore expect that antisite defects are the dominant intrinsic defects in Fe$_2$Nb. This is consistent with the experimentally observed absence of vacancies in C14-Fe$_2$Nb~\cite{Zhu1999}.

The energetic ordering of the antisite defects in Tab.~\ref{tab:ef_vac_anti} is related to the size of the Fe/Nb atoms on one hand and the Z12/Z16 coordination polyhedra on the other hand: The DFT results indicate that it is energetically favorable by about at least 0.5~eV to squeeze a comparably larger Nb atom in an empty Z12 polyhedron of a removed Fe atom than to fill an empty Z16 polyhedron of a removed Nb atom with a comparably smaller Fe atom.

\subsubsection{Influence of point defects on generalized SFE}

An early work on C14-Fe$_2$Nb reported that the antisite defects cause a hardening on both sides of the 2:1 stoichiometry~\cite{Zhu1999}. 
More recent works made contrasting observations of progressive softening for off-stoichiometric compositions~\cite{Voss2009,Takata2016}. 
This softening was suggested to arise from the free volume generated by the vacancy or antisite which facilitates the motion of synchro-Shockley dislocations~\cite{Kumar2004,Chu1994}. 
In order to shed light on these contrasting findings, we compute the influence of vacancy and antisite defects on the barrier for plastic deformation in C14-Fe$_2$Nb.
In particular, we compute the formation energy of the unstable and stable SF ($x=0.5$ and $x=1.0$ in Fig.~\ref{fig:SFE}) with either of these two point defects present in different distances from the SF plane. 

\begin{figure}[htb]
 \centering
 \subfigure[antisite]{
 \includegraphics[width=0.95\columnwidth]{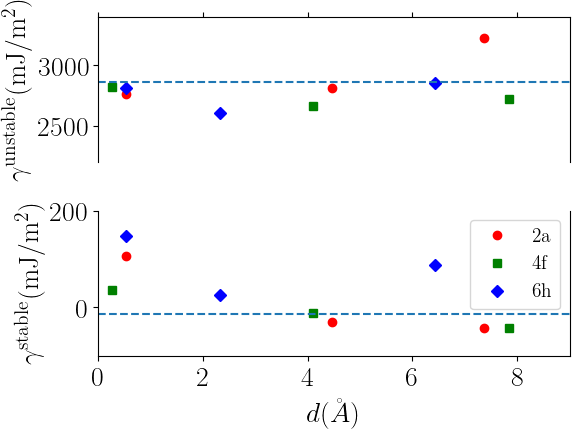}}
 \subfigure[vacancy]{
 \includegraphics[width=0.95\columnwidth]{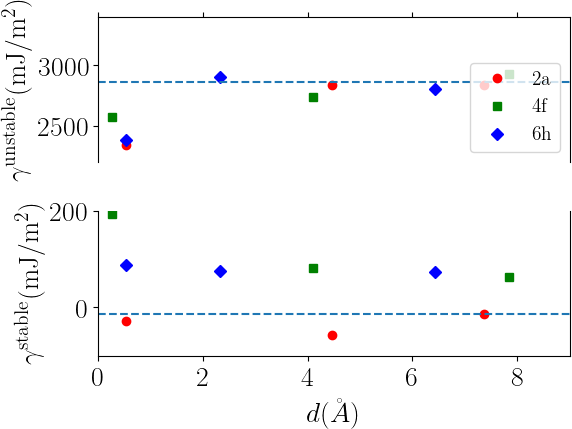}}
 \caption{Influence of (a) antisite defect and (b) vacancy on the energy of stable and unstable stacking fault in C14-Fe$_2$Nb as a function of the distance from the respective stacking fault plane.}
 \label{fig:sf_defect}         
\end{figure}

Our calculations, compiled in Fig.~\ref{fig:sf_defect}, show that nearly all defects increase the formation energy of the stable stacking fault ($\gamma^{\mathrm stable}$) and decrease the energy barrier for synchroshear ($\gamma^{\mathrm unstable}-\gamma^{\mathrm stable}$).
The only exception with a notably increased barrier is the antisite defect on $2a$ Fe sites in 7.5\AA\, distance from the SF plane.
The origin of this exception is the convergence to a magnetic state that is different from the one of the unsheared structure.  
The variation of the SF energy with distance for the individual defects shows that only vacancies on $2a$ Fe sites are weakly attracted to the SF plane while all other vacancy and antisite defects are weakly repelled. 
From the overall rather small thermodynamic driving forces for segregation we expect that the local site occupation is not considerably affected by the presence of the stable SF.

In summary, our calculations show that deviations from the 2:1 composition due to vacancies or due to antisite defects lead to a softening of C14-Fe$_2$Nb on both sides of the stoichiometry due to a reduced energy barrier for shearing.
This finding explains the recent experimental observations~\cite{Voss2009,Takata2016} of softening for off-stoichiometric compositions.

\subsection{Polytypic transformations}
\label{sec:polytypic}

\begin{figure}[htb]
 \centering
  \subfigure[C14$\rightarrow$C15]{\label{fig:trans_C14_C15}
 \includegraphics[width=0.95\columnwidth]{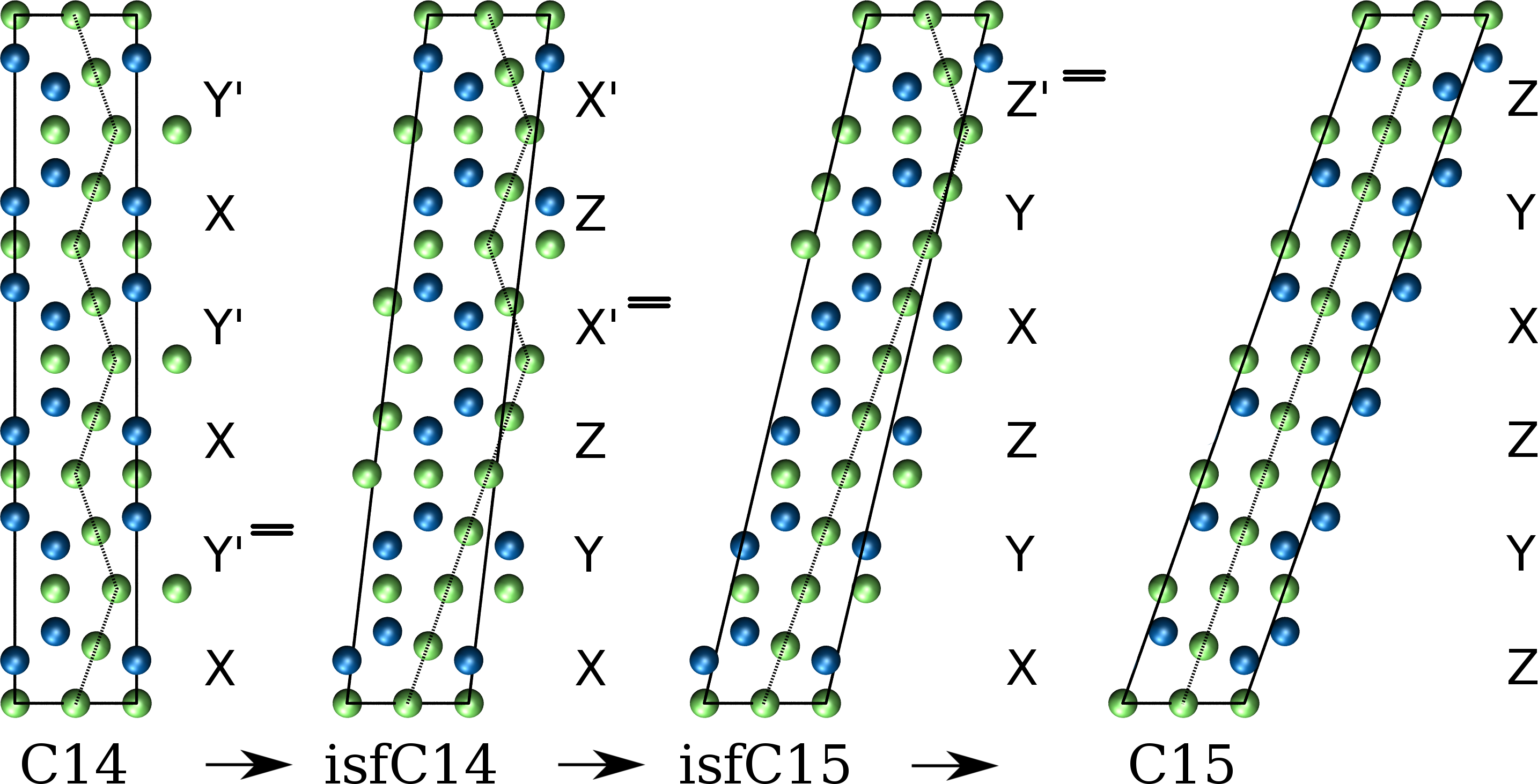}}
  \subfigure[C14$\rightarrow$C36$\rightarrow$C15]{\label{fig:trans_C14_C36_C15}
 \includegraphics[width=0.95\columnwidth]{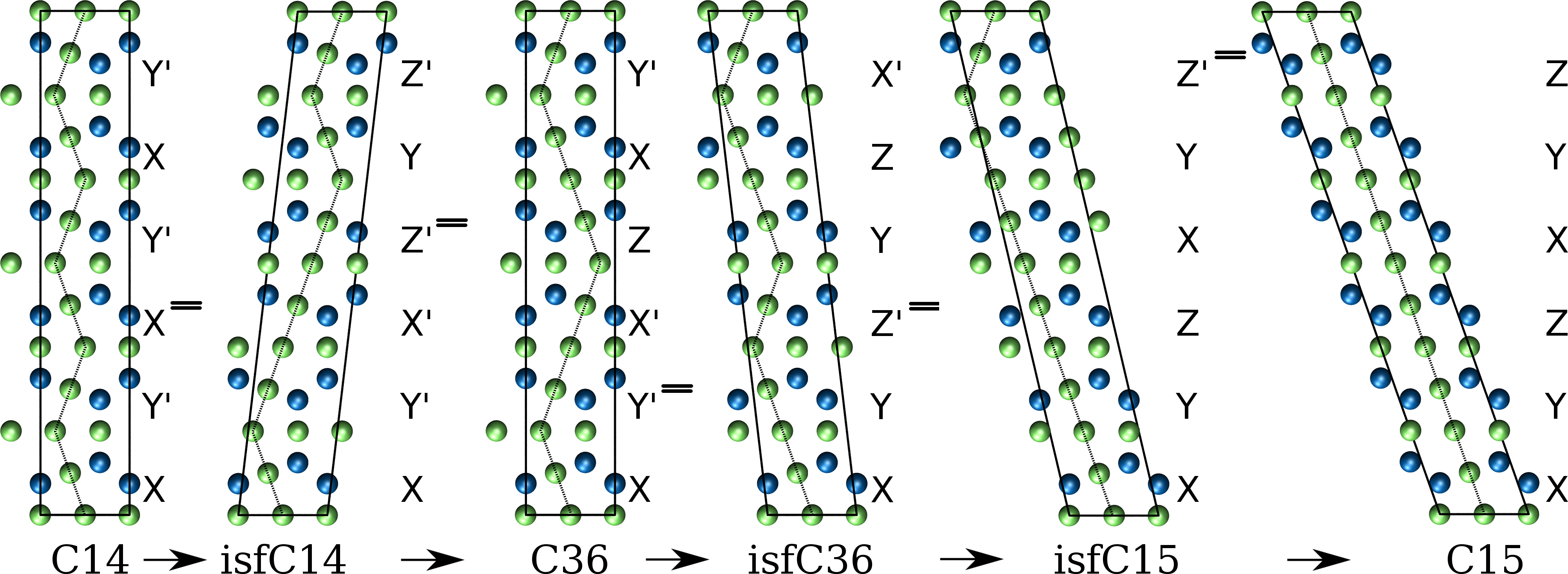}}
 \caption{Series of synchroshear deformations of the polytypic transformations routes from C14 to C15 with intermediate intrinsic stacking fault (isf) configurations. The synchroshear plane and the characteristic layer configuration are indicated.}
 \label{fig:trans}         
\end{figure}

The stresses that are needed to overcome the computed barriers for basal slip (Sec.~\ref{sec:SFE}) are so large that one might expect that other processes are initiated, too.
A particularly likely candidate are stress-induced transformations that are known from other Laves phase compounds~\cite{Stein2005}.
Here, we compute the energy barriers of polytypic transformation between C14, C15 and C36 Laves phases by successive synchroshear, see Fig.~\ref{fig:trans}.

A synchroshear deformation can be seen as propagation of two Shockley dislocations through the quadruple layer which leads to a macroscopic shear strain. 
The Burgers vector of the Shockley dislocations can be one of those shown in Fig.~\ref{fig:laves_sites}. 
If the layer of small atoms in the triple layer moves along $-b_1$, the upper layer of large atoms must move in synch either along $b_2$ or
$b_3$.
This deformation transforms a quadruple layer $t$ to $t'$ or vice versa.
Applying an equal number of these three equivalent Burgers vectors can lead to an alternative polytypic transformation without macroscopic strain~\cite{Kumar2004}. 
A transformation between Laves phase polytypes can also be realized if the number of $t\rightarrow t'$ and $t'\rightarrow t$ synchroshear steps are equal.
This is the case for the C14$\rightarrow$C36 transformation shown in Fig.\ref{fig:trans_C14_C36_C15}.
Therefore, it is possible to transform the C14 structure to the C15 structure in two ways~\cite{Vedmedenko2008} by a series of synchroshear steps as illustrated in Fig.~\ref{fig:trans}.
In the plastic deformation of C14-Fe$_2$Nb, an already formed stable SF could transform further by successive synchroshear steps that eventually lead to a precipitate of C15 or C36.

We compare the formation of a new stable SF to a further synchroshear of an existing SF, in terms of the energy profile along the polytypic transformation by successive synchroshear.
We used unrelaxed atomic structures along the transformation path that we compare to the unrelaxed structures of the generalized SFE for basal slip in Fig.~\ref{fig:SFE}.
From our calculations for the SFE we expect that including the structural relaxation would lead to an overall scaling that would not alter our conclusions.

\begin{figure}[htb]
 \centering
 \subfigure[C14$\rightarrow$C15]{ 
 \includegraphics[width=0.95\columnwidth]{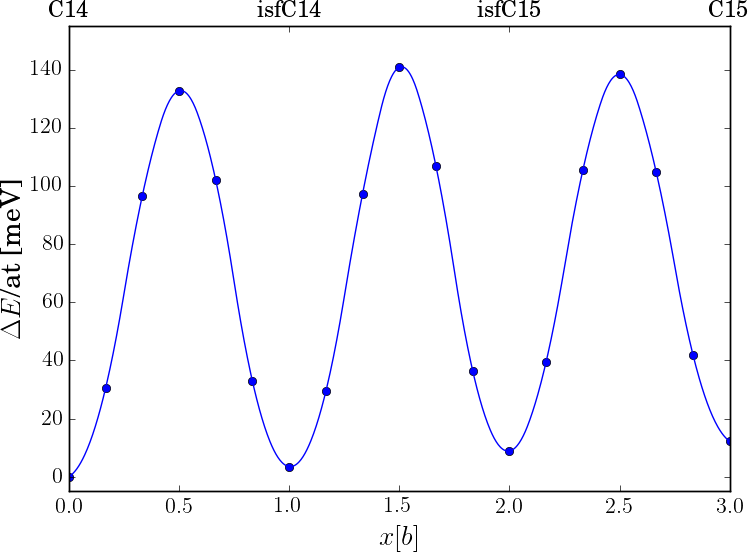}
 \label{fig:ene_t1}         
 }
 \subfigure[C14$\rightarrow$C36$\rightarrow$C15]{
 \centering
 \includegraphics[width=0.95\columnwidth]{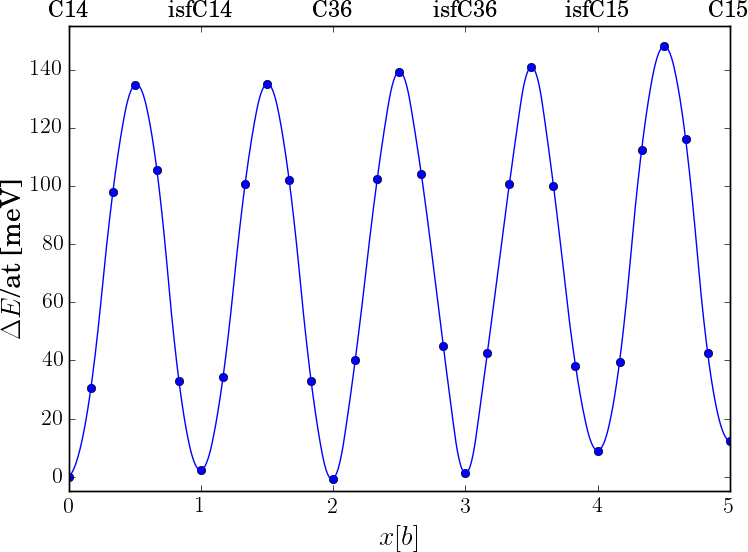}
 \label{fig:ene_t2}         
 }
 \caption{Change in energy along polytypic transformations from C14 to C15 and C36. The points denote the DFT calculations; the lines are interpolations to guide the eye.}
 \label{fig:ene_t}         
\end{figure}

In Fig.~\ref{fig:ene_t} we plot the computed change in energy with respect to the initial C14-Fe$_2$Nb for the corresponding configurations.
The first insight from our DFT results is that all intermediate stable states of the polytypic transformation, i.e. the Laves phases as well as their intrinsic stacking faults, are very close in energy within a range of less than 20~meV per atom.
The formation energy of the SF in C36 is negative since C36 is predicted to be more energetically favorable in our spin-polarized DFT calculations.
The correct prediction of the C14 ground state requires to extend the DFT calculations to a proper treatment of paramagnetism~\cite{Slapakova-2020} which would then be too computationally involved for the large supercells required for this work.
The second insight is that the energy barriers between the intermediate stable states of the synchroshear series are almost identical ($\approx$ 140 meV/at $=$ 4~J/m$^2$). 
Moreover, they are very similar to the energy barrier for a single synchroshear or undulating slip in the generalized SFE curve of C14-Fe$_2$Nb for unrelaxed atomic positions (Fig.\ref{fig:SFE}).
This indicates that pressures which lead to the formation of an intrinsic SF in C14-Fe$_2$Nb may also initiate a polytypic transformation of C14 to C15 or C36 by further transforming already existing SFs.

\begin{figure}[htb]
 \centering
 \subfigure[C14$\rightarrow$C15]{
 \includegraphics[width=0.95\columnwidth]{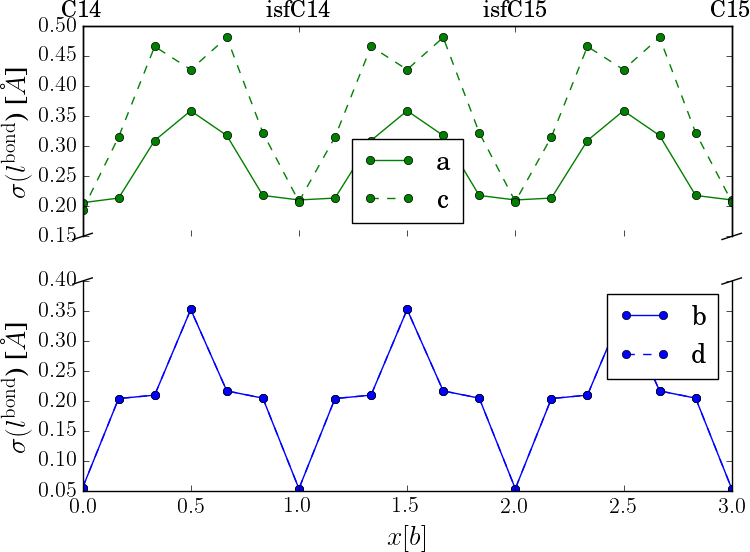}
 \label{fig:ana_C14_C15}         
 }
 \subfigure[C14$\rightarrow$C36$\rightarrow$C15]{

 \includegraphics[width=0.95\columnwidth]{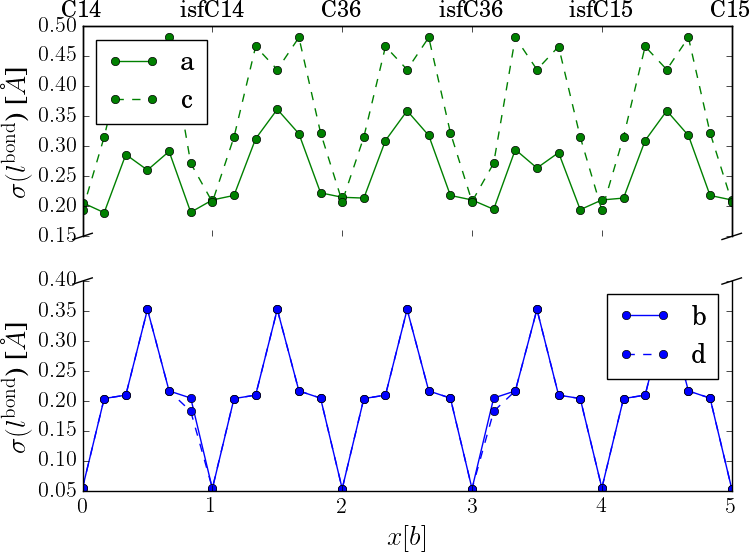}
 \label{fig:ana_C14_C36_C15}         
}
 \caption{Bond lengths in the coordination polyhedra of atoms comprising the synchrosheared quadruple layer along the transformation path:
         (a) C14$\rightarrow$C15 (b) C14$\rightarrow$C36$\rightarrow$C15. 
         The labels a, b, c and d correspond to the atoms in the stacking sequence S-L-S-L in Fig.~\ref{fig:laves_quadruple}.  
         }
\label{fig:ana_trans}
\end{figure}

The energy profile along the polytypic transformation by a series of synchroshear steps can be rationalized in terms of the corresponding deformation of the Z12 and Z16 Frank Kasper polyhedra.
A measure of this deformation is the change in the nearest-neighbour bond-lengths, $\sigma$, of the S and L atoms in the synchrosheared quadruple layer shown in Fig.~\ref{fig:ana_trans}. 
The change in bond lengths along the synchroshear steps is similar for the S and L atoms in in the quadruple layer and reaches values of nearly 0.5\,\AA.
This considerable change is directly related to the energy increase at the transition point of the energy profile in Fig.~\ref{fig:ene_t}.

\section{Conclusions}
\label{sec:conclusions}

We perform DFT calculations on the role of different basal slip mechanisms for the plastic deformations of the C14-Fe$_2$Nb Laves phase.
The generalized stacking-fault energy curves show that synchroshear and undulating slip pass the same transition state with an energy barrier of 3~$J/m^2$ that is considerably lower than direct slip.
The experimentally observed stacking fault configurations are in line with the final structures of both, synchroshear and undulating slip.
However, the large energy barriers for the considered mechanisms and the rather low formation energy of stable SF in C14 may indicate the existence of not yet considered mechanisms that lead to the experimentally observed SF configurations.

Our DFT calculations for the stable and unstable SF in the presence of vacancies and antisite defects show that the barrier for synchroshear is decreased.
This finding explains the experimentally observed softening of C14-Fe$_2$Nb by off-stoichoimetric compositions on both sides of the 2:1 composition.

In order to assess the possibility of polytypic transformations, we compute the energy profile of the transformation from C15 to C14 and C36 by a series of synchroshear deformations. 
The energy barriers are very similar for the different steps of the deformation series and moreover very similar to synchroshear and undulating slip. 
This indicates that the pressures which lead to plastic deformation of C14-Fe$_2$Nb by stacking fault formation can also initiate polytypic transformation of C14 to C15 or C36 through successive synchroshear of existing SF.

\section*{Appendix: Formation energy of point defects}

For the energetic ranking of the different point defects in C14-Fe$_2$Nb, we compute the formation energy per atom of a general structure with $N$ Fe atoms and $M$ Nb atoms as
\begin{equation}\label{eq:dH}
    \Delta H_{\mathrm{struc}}^{N,M} = \frac{E_{\mathrm{struc}}^{N,M} - N E_{\mathrm{Fe}} - M E_{\mathrm{Nb}}}{N+M}
\end{equation}
where $E_{\mathrm{struc}}^{N,M}$, $E_{\mathrm{Fe}}$ and $E_{\mathrm{Nb}}$ are the DFT total energies of the respective general structure and the energetically most favorable bcc structure of Fe and Nb. This choice corresponds to the chemical potential of the DFT convex hull at T=0K. The superscripts denote the number of Fe and Nb atoms in the DFT simulation cell. The bcc simulation cells are assumed to contain one atom.
The formation energy per atom of ideal C14-Fe$_2$Nb without defects is given by 
\begin{equation}\label{eq:dHC14}
    \Delta H_{C14}^{N,M} = \frac{E_{C14}^{N,M} - N E_{\mathrm{Fe}} - M E_{\mathrm{Nb}}}{N+M}
\end{equation}
and referred to as $\Delta H_{\mathrm{ideal}}$ in Eqs.~\ref{eq:dH1} and~\ref{eq:dH2}. The formation of vacancies and antisite defects in C14-Fe$_2$Nb alters the chemical composition. In order to compute the defect formation energy, we need the energy difference between ideal bulk and the defect structure plus additional terms that balance the total number of atoms for each species. 

For vacancies, the formation energy per atom of a supercell structure with a missing Fe atom ('C14vacFe') is given by 
\begin{equation}\label{eq:dHvac}
    \Delta H_{\mathrm{C14vacFe}}^{N-1,M} = \frac{E_{\mathrm{C14vacFe}}^{N-1,M} - (N-1) E_{\mathrm{Fe}} - M E_{\mathrm{Nb}}}{N-1+M} \, .
\end{equation}
The formation energy per vacancy
\begin{equation}
    E_{f}^{vac} = E_{\mathrm{C14vacFe}}^{N-1,M} - E_{C14}^{N,M} + E_{\mathrm{Fe}}  
\end{equation}
can be rewritten in terms of the structure formation energy (per atom) as
\begin{eqnarray}
    E_{f}^{vac} &=& E_{\mathrm{C14vacFe}}^{N-1,M} - (N-1) E_{\mathrm{Fe}} - M E_{\mathrm{Nb}}\nonumber \\
                && - E_{C14}^{N,M} + (N-1) E_{\mathrm{Fe}} + M E_{\mathrm{Nb}}\nonumber \\
                && + E_{\mathrm{Fe}}  
\end{eqnarray}
which by comparison to Eqs.~\ref{eq:dHC14} and~\ref{eq:dHvac} reduces to
\begin{eqnarray}
    E_{f}^{vac} = (N-1+M) \Delta H_{\mathrm{C14vacFe}}^{N-1,M} - (N+M) \Delta H_{C14}^{N,M} \, .
\end{eqnarray}
With the fraction of the number of defects to the number of atoms in the defect-free supercell of $x_D=1/(N+M)$ we obtain
\begin{eqnarray}
    E_{f}^{vac} = \frac{ (1-x_D) \Delta H_{\mathrm{C14vacFe}}^{N-1,M} - \Delta H_{C14}^{N,M}}{x_D} 
\end{eqnarray}
as formation energy of an Fe vacancy, and correspondingly for a Nb vacancy. 

For an antisite defect, e.g. an Fe atom on a Nb site ('C14antiFe'), the formation energy per antisite defect is
\begin{equation}
    E_{f}^{\mathrm{antiFe}} = E_{\mathrm{C14antiFe}}^{N+1,M-1} - E_{C14}^{N,M} - E_{\mathrm{Fe}} + E_{\mathrm{Nb}}
\end{equation}
which can be rewritten as
\begin{eqnarray}
    E_{f}^{\mathrm{antiFe}} &=& E_{\mathrm{C14antiFe}}^{N+1,M-1} - (N+1) E_{\mathrm{Fe}} - (M-1) E_{Nb}\nonumber \\
                && - E_{C14}^{N,M} + (N+1) E_{\mathrm{Fe}} + (M-1) E_{Nb}\nonumber \\
                &&  - E_{\mathrm{Fe}} + E_{\mathrm{Nb}} \, .
\end{eqnarray}
Following the procedure above we obtain
\begin{eqnarray}
    E_{f}^{\mathrm{antiFe}} = (N+M) \Delta H_{\mathrm{C14antiFe}}^{N+1,M-1} - (N+M) \Delta H_{C14}^{N,M}
\end{eqnarray}
that reduces to
\begin{eqnarray}
    E_{f}^{\mathrm{antiFe}} = \frac{ \Delta H_{\mathrm{C14antiFe}}^{N+1,M-1} - \Delta H_{C14}^{N,M}}{x_D} 
\end{eqnarray}
and correspondingly for Nb atoms on Fe sites.

\section*{Acknowledgements}
We acknowledge fruitful discussions with Ali Zendegani and Tilmann Hickel.
This work was funded by the German Research Society DFG (project number 289654611).

\bibliographystyle{elsarticle-num}
\bibliography{fenb_sf}

\begin{thebibliography}{10}
\expandafter\ifx\csname url\endcsname\relax
  \def\url#1{\texttt{#1}}\fi
\expandafter\ifx\csname urlprefix\endcsname\relax\def\urlprefix{URL }\fi
\expandafter\ifx\csname href\endcsname\relax
  \def\href#1#2{#2} \def\path#1{#1}\fi

\bibitem{Sinha1978}
A.~Sinha, Topologically close-packed structures of transition metal alloys,
  Progress in Materials Science 15 (1972) 81.

\bibitem{Stein2021}
F.~Stein, A.~Leineweber, Laves phases: a review of their functional and
  structural applications and an improved fundamental understanding of
  stability and properties, Journal of Materials Science 56 (2021) 5321.

\bibitem{Seiser-2011-1}
B.~Seiser, R.~Drautz, D.~G. Pettifor, {TCP} phase predictions in {Ni}-based
  superalloys: Structure maps revisited, Acta materialia 59 (2011) 749.

\bibitem{Seiser-2011-2}
B.~Seiser, T.~Hammerschmidt, A.~N. Kolmogorov, R.~Drautz, D.~G. Pettifor,
  Theory of structural trends within 4d and 5d transition metals topologically
  close-packed phases, Physical Review B 83 (2011) 224116.

\bibitem{Hammerschmidt2013}
T.~Hammerschmidt, A.~Bialon, D.~Pettifor, R.~Drautz, Topologically close-packed
  phases in binary transition-metal compounds: matching high-throughput ab
  initio calculations to an empirical structure map, New Journal of Physics 15
  (2013) 115016.

\bibitem{Ladines2015}
A.~Ladines, T.~Hammerschmidt, R.~Drautz, Structural stability of {Fe}-based
  topologically close-packed phases, Intermetallics 59 (2015) 59.

\bibitem{Livingston1992}
J.~D. Livingston, Laves-phase superalloys?, physica status solidi (a) 131
  (1992) 415.

\bibitem{Heggen2010}
M.~Heggen, L.~Houben, M.~Feuerbacher, Plastic deformation mechanism in complex
  solids, Nature Materials 9 (2010) 332--336.

\bibitem{Zhang2020}
Y.~Zhang, W.~Zhang, B.~Du, W.~Li, L.~Sheng, H.~Ye, K.~Du, Shuffle and glide
  mechanisms of prismatic dislocations in a hexagonal $c14$-type laves-phase
  intermetallic compound, Phys. Rev. B 102 (2020) 134117.

\bibitem{Kumar2004}
K.~Kumar, P.~Hazzledine, Polytypic transformations in {Laves} phases,
  Intermetallics 12 (2004) 763.

\bibitem{Allen1972}
C.~W. Allen, P.~Delavignette, S.~Amelinckx, Electron microscopic studies of the
  {Laves} phases {Ti}{Cr}$_2$ and {Ti}{Co}$_2$, physica status solidi (a) 9
  (1972) 237.

\bibitem{Kumar1994}
K.~Kumar, D.~Miracle, Microstructural evolution and mechanical properties of a
  {Cr}-{Cr}$_2${Hf} alloy, Intermetallics 2 (1994) 257.

\bibitem{Chisholm2005}
M.~F. Chisholm, S.~Kumar, P.~Hazzledine, Dislocations in complex materials,
  Science 307 (2005) 701--703.

\bibitem{Hazzledine1993}
P.~Hazzledine, P.~Pirouz, Synchroshear transformations in {Laves} phases,
  Scripta Metallurgica et Materialia 28 (1993) 1277 -- 1282.

\bibitem{Zhang2011}
W.~Zhang, R.~Yu, K.~Du, Z.~Cheng, H.~Zhu, J.and~Ye, Undulating slip in {Laves}
  phase and implications for deformation in brittle materials, Physical Review
  Letters 106 (2011) 165505.

\bibitem{Tadmor2003}
E.~Tadmor, S.~Hai, A peierls criterion for the onset of deformation twinning at
  a crack tip, Journal of the Mechanics and Physics of Solids 51 (2003) 765.

\bibitem{VanSwygenhoven2004}
H.~Van~Swygenhoven, P.~M. Derlet, A.~G. Froseth, Stacking fault energies and
  slip in nanocrystalline metals, Nature Materials 3 (2004) 399.

\bibitem{Moeller2018}
J.~M\"oller, M.~Mrovec, I.~Bleskov, T.~Hammerschmidt, R.~Drautz,
  C.~Els{\"a}sser, J.~Neugebauer, T.~Hickel, E.~Bitzek, On {110} planar faults
  in strained bcc metals - origins and implications of a commonly observed
  artefact of classical potentials, Physical Review Materials 2 (2018) 093606.

\bibitem{Vedmedenko2008}
O.~Vedmedenko, F.~R{\"o}sch, C.~Els{\"a}sser, First-principles density
  functional theory study of phase transformations in {Nb}{Cr}$_2$ and
  {Ta}{Cr}$_2$, Acta Materialia 56 (2008) 4984 -- 4992.

\bibitem{Ma2013}
L.~Ma, T.-W. Fan, B.-Y. Tang, L.-M. Peng, W.-J. Ding, Ab initio study of {I2}
  and {T2} stacking faults in {C14} {Laves} phase {Mg}{Zn}$_2$, The European
  Physical Journal B 86 (2013) 188.

\bibitem{Ma2014}
L.~Ma, R.-K. Pan, S.-C. Zhou, T.-P. Luo, D.-H. Wu, T.-W. Fan, B.-Y. Tang, Ab
  initio study of stacking faults and deformation mechanism in {C15} {Laves}
  phases {Cr}$_2${X} ({X = Nb, Zr, Hf}), Materials Chemistry and Physics 143
  (2014) 702.

\bibitem{Voss-11}
S.~Vo\ss, M.~Palm, F.~Stein, D.~Raabe, Phase equilibria in the {Fe}-{Nb}
  system, Journal of Phase Equilibria and Diffusion 32 (2011) 97--104.

\bibitem{Kresse1996}
G.~Kresse, J.~Furthm{\"u}ller, Efficiency of ab-initio total energy
  calculations for metals and semiconductors using a plane-wave basis set,
  Computational Materials Science 6 (1996) 15.

\bibitem{Kresse1996b}
G.~Kresse, J.~Furthm\"uller, Efficient iterative schemes for ab initio
  total-energy calculations using a plane-wave basis set, Physical Review B 54
  (1996) 11169.

\bibitem{Kresse1999}
G.~Kresse, D.~Joubert, From ultrasoft pseudopotentials to the projector
  augmented-wave method, Physical Review B 59 (1999) 1758.

\bibitem{Bloechl-94}
P.~E. Bl\"ochl, Projector augmented-wave method, Physical Review B 50 (1994)
  17953.

\bibitem{Perdew-96}
J.~P. Perdew, K.~Burke, M.~Ernzerhof, Generalized gradient approximation made
  simple, Physical Review Letters 77 (1996) 3865.

\bibitem{Takata2008}
N.~Takata, S.~Ishikawa, T.~Matsuo, M.~Takeyama, Transmission electron
  microscopy of {Fe}$_2${Nb} {Laves} phase with {C14} structure in {Fe-Nb-Ni}
  alloys, MRS Online Proceedings Library 1128 (2008) 806.

\bibitem{Slapakova2016}
M.~Slapakova~Pokova, S.~Vo\ss, K.~Kumar, C.~Liebscher, F.~Stein, Transmission
  electron microscopy of deformed {Laves} phase {Nb}{Fe}$_2$, in: European
  Microscopy Congress 2016 Proceedings, Wiley-VCH, 2016, pp. 263--264.

\bibitem{Guenole-19}
J.~Guenole, F.-Z. Mouhiba, L.~Huber, B.~Grabowski, S.~Korte-Kerzel, Basal slip
  in {Laves} phases: The synchroshear dislocation, Scripta Materialia 166
  (2019) 134.

\bibitem{Thoma1995}
D.~Thoma, J.~Perepezko, A geometric analysis of solubility ranges in {Laves}
  phases, Journal of Alloys and Compounds 224 (1995) 330.

\bibitem{Liu2000}
C.~Liu, J.~Zhu, M.~Brady, C.~McKamey, L.~Pike, Physical metallurgy and
  mechanical properties of transition-metal {Laves} phase alloys,
  Intermetallics 8 (2000) 1119.

\bibitem{Ladines-17}
A.~Ladines, R.~Drautz, T.~Hammerschmidt, Ab-initio study of {C} and {N} point
  defects in the {C14}-{Fe}$_2${Nb} phase, Journal of Alloys and Compounds 693
  (2017) 1315--1322.

\bibitem{Zhu1999}
J.~Zhu, L.~Pike, C.~Liu, P.~Liaw, Point defects in binary {Laves} phase alloys,
  Acta Materialia 47~(7) (1999) 2003 -- 2018.

\bibitem{Voss2009}
S.~Vo\ss, F.~Stein, M.~Palm, D.~Gr{\"u}ner, G.~Kreiner, G.~Frommeyer, D.~Raabe,
  Composition dependence of the hardness of {Laves} phases in the {Fe}-{Nb} and
  {Co}-{Nb} systems, MRS Online Proceedings Library 1128 (2008) 805.

\bibitem{Takata2016}
N.~Takata, H.~Ghassemi-Armaki, M.~Takeyama, S.~Kumar, Nanoindentation study on
  solid solution softening of {Fe}-rich {Fe}$_2${Nb} {Laves} phase by {Ni} in
  {Fe-Nb-Ni} ternary alloys, Intermetallics 70 (2016) 7 -- 16.

\bibitem{Chu1994}
F.~Chu, D.~Pope, Deformation of {C15} laves phase alloys, Materials Research
  Society Symposium Proceedings 364 (1994) 1197.

\bibitem{Stein2005}
F.~Stein, M.~Palm, G.~Sauthoff, Structure and stability of {Laves} phases part
  {II} - {S}tructure type variations in binary and ternary systems,
  Intermetallics 13 (2005) 1056.

\bibitem{Slapakova-2020}
M.~Slapakova, A.~Zendegani, C.~Liebscher, T.~Hickel, J.~Neugebauer,
  T.~Hammerschmidt, A.~Ormeci, Y.~Grin, G.~Dehm, S.~Kumar, F.~Stein, Atomic
  scale configuration of planar defects in the {Nb}-rich {C14} {L}aves phase
  {NbFe}$_2$, Acta Mater. 183 (2020) 362--376.

\end{thebibliography}

\end{document}